\documentclass[twocolumn,prl,epsfig,aps,showpacs]{revtex4-1}

\usepackage{graphicx}
\usepackage{color}

\newcommand{\beq}{\begin{eqnarray}}
\newcommand{\eeq}{\end{eqnarray}}

\begin{document}
%\date{\today}
\setcounter{page}{1}

\title{Liquid and solid phases of $^3$He on graphite 
}
\author{M. C. Gordillo}
\affiliation{Departamento de Sistemas F\'{\i}sicos, Qu\'{\i}micos y Naturales,
Universidad Pablo de Olavide. E-41013 Seville, Spain}
\author{J. Boronat}
\affiliation{Departament de F\'{\i}sica, 
Universitat Polit\`ecnica de Catalunya, 
Campus Nord B4-B5, E-08034 Barcelona, Spain}

\begin{abstract}
Recent heat-capacity experiments show quite unambiguously the existence of
a liquid $^3$He phase adsorbed on graphite. This liquid is stable at an
extremely low density, possibly one of the lowest found in Nature. Previous
theoretical calculations of the same system, and in strictly two
dimensions, agree with the result that this
liquid phase is not stable and the system is in the gas phase.   
We calculated the phase diagram of normal $^3$He adsorbed on graphite at
$T=0$ using quantum Monte Carlo methods.
Considering a  
fully corrugated
substrate we observe that at densities lower that 0.006 \AA$^{-2}$ 
the system is a very dilute gas, that at that density  is in
equilibrium with a liquid of density 0.014  \AA$^{-2}$. 
Our prediction matches very well the recent experimental findings on the
same system.
On the contrary, 
when a flat substrate is considered, no gas-liquid coexistence
is found, in agreement with previous calculations.   
We also report results on the different solid structures, and the corresponding
phase transitions that appear at higher densities.
\end{abstract}


\maketitle

Recent heat capacity measurements of $^3$He adsorbed on graphite by
Sato \textit{et al.} \cite{fukuyama,fukuyama1} have shown that its monolayer is a stable liquid in
the density range $0.006$-$0.009$ \AA$^{-2}$. 
One of the most interesting
aspects of this phase is its extremely low density, with interparticle
distances as large as 10 \AA, which could constitute one of the
lowest-density stable liquids in nature.    
This new
finding has re-opened an old issue that has been under discussion for more than
thirty years, i.e., the nature (gas or liquid) of two-dimensional (2D)
$^3$He~\cite{gasparinip,smart}. Previous experiments showed contradictory
results due in part to the different setups and employed 
substrates~\cite{gasparini,godfrin,greywall,hallock}. Now,
the new data from Ref. \cite{fukuyama} on a clean graphite substrate
seem to incline the debate towards the confirmation of
this liquid phase existence. 
%Moreover, in this experiment the same
%result is obtained for the first monolayer on graphite and for the
%monolayer on graphite preplated with $^4$He. 

On the theoretical side, there is a broad consensus on the gas
character of strictly 2D $^3$He~\cite{novaco,miller,hnc,chester,grau}.
However, the practical need of a substrate to actually realize the $^3$He monolayer
could modify this result. Previous attempts to calculate the
properties of the adsorbed monolayer in a strongly attractive substrate
such as graphite arrived to the same result. In Ref. \cite{bonin}, it is
shown that the possibility of $^3$He atoms moving  perpendicularly to
the surface leads to a stable liquid phase when the substrate is weakly
attractive, as on some alkali metal surfaces. This is probably expected
because the system goes from a 2D film to a three-dimensional (3D) configuration where
liquid $^3$He is the ground-state phase. 
 
In this work, we concerned ourselves with the adsorption of $^3$He on a
clean surface of graphite, 
trying to reproduce the recent experimental findings of Sato
\textit{et al.}~ \cite{fukuyama} 
Our goal was to bridge the    discrepancy
between the strictly 2D calculations and the experimental data by improving the
theoretical description of the system. Since considering a quasi-two
dimensional flat adsorbent is clearly  not enough for graphite \cite{bonin}, we included
the effects of the substrate corrugation  on the 
behavior of the adsorbate,
in line with what has been done previously for $^4$He on the same system 
\cite{yo,yo3,yo2,manu1,manu2,manu3,manu4,manu5}. We found that a corrugated surface is the missing
ingredient to reconcile the experimental and theoretical data. In
addition, this approach allows us also to calculate the entire phase
diagram of $^3$He on  graphite, including the commensurate solids that
cannot appear in a strictly 2D model.  

Given the low temperatures involved in the experiments (of the order of
mK), it is reasonable to think that the ground state of $^3$He on graphite
is a reasonable description of the system under consideration. To obtain
it, we have to solve the Schr\"odinger equation corresponding to 
the
many-body Hamiltonian, 
\begin{equation} \label{hamiltonian} H =
\sum_{i=1}^N  \left[ -\frac{\hbar^2}{2m} \nabla^2 + V_{\text{ext}}(x_i,y_i,z_i)
\right] + \sum_{i<j}^N V_{\text{He-He}} (r_{ij}), 
\end{equation} 
where $x_i,y_i$
and $z_i$ are the coordinates for each of the $N$ $^3$He atoms, and $m$
their mass. Following Ref. \cite{yo},  graphite was modeled by a set
of eight graphene layers separated $3.35$ \AA$ $  in the $z$ direction and
stacked in the A-B-A-B way typical of this compound. All the individual
carbon atoms in each layer were considered.  $V_{\text{ext}}(x_i,y_i,z_i)$ was the
sum of all the C-He atomic interactions, calculated using the Carlos  and
Cole anisotropic potential \cite{carlosandcole}, which has been widely
used in calculations of $^4$He adsorbed on graphite. $V_{\text{He-He}} (r)$ is
the standard Aziz potential \cite{aziz}, that depends on the distance
$r_{ij}$ between $^3$He atoms.  

\begin{figure}
\begin{center}
\includegraphics[width=0.7\linewidth]{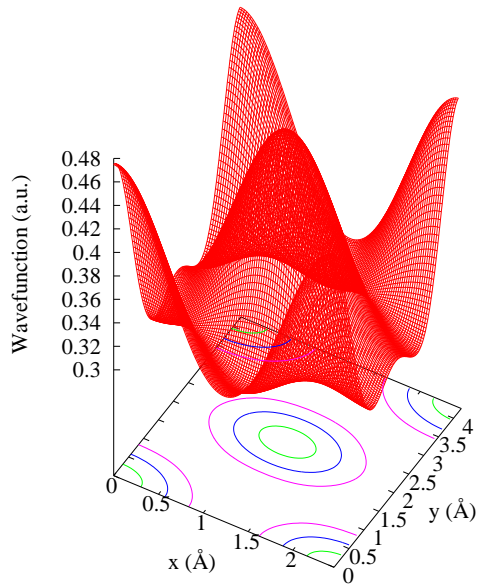} 
\caption{(Color online)  Plot of $u(x,y,z=2.88)_{(0,0)}$ with its corresponding 
contour map showing the corrugation of the one-particle part of the trial
wavefunction.      
}
\label{cut}
\end{center}
\end{figure}

To solve the Schr\"odinger equation describing the system, we used the
diffusion Monte Carlo (DMC) method. For a set of bosons, DMC allows us
to obtain exactly the energy of their  ground state, within the statistical
uncertainties derived from the stochastic nature of the method. 
However, when we deal
with fermions, as in the present case, the sign problem makes an exact
calculation not possible. We follow the usual approach in which one imposes
that the nodal surface is the one of the trial wavefunction used as
guiding function in the DMC algorithm~\cite{hammond}. This approximation is known as
fixed-node method (FN) and provides an upper bound to the exact
ground-state energy of the system. 
We chose as a  trial wavefunction
\begin{equation}
\Phi({\bf r_1, r_2, ..., r_N}) =  D^{\uparrow} D^{\downarrow}  \prod_{i<j} \exp \left[-\frac{1}{2}
\left(\frac{b_{\rm He-He}}{r_{ij}} \right)^5 \right] 
\label{trial1}
\end{equation}
where ${\bf r_1, r_2, ..., r_N}$ are the helium coordinates. 
The parameter $b_{\rm He-He}$ in the Jastrow part of Eq.
(\ref{trial1}) was taken to be  $2.96$ \AA, as in a purely two-dimensional
system \cite{grau}.    $D^{\uparrow}$ and $D^{\downarrow}$ are Slater
determinants that depend on the coordinates of the spin-up and spin-down
atoms, respectively. We considered  always an
unpolarized phase, $N_{\uparrow} =
N_{\downarrow} = N/2$. The single-particle functions entering those determinants,
$\psi({\bf r_i})$, were the solutions of the Schr\"odinger equation derived
from the one-body Hamiltonian resulting  of dropping the interparticle
interaction [last term in Eq.
(\ref{hamiltonian})].  Since $V_{\text{ext}}(x_i,y_i,z_i)$ has the periodicity of
the underlying substrate, we can invoke Bloch's theorem to
write~\cite{ashcroft} 
\begin{equation} \label{Bloch}  \psi({\bf r_i}) = u({\bf
r_i})_{\bf k} \exp(i k_x x_i + i k_y y_i), 
\end{equation} 
where $u({\bf
r_i})_{\bf k}$ obeys the Born-von Karman periodic boundary conditions (in
2D) with respect to the unit cell whose replication defines the
graphite structure. We chose as unit   cell one whose surface is 2.46
$\times$ 4.26 \AA$^2$, that includes four carbon atoms in its upper layer
(the ones that form  the characteristic  hexagon
of a honeycomb
arrangement). In general, $u({\bf r_i})_{\bf k}$  depends on the reciprocal
vector {\bf k} = $(k_x,k_y)$. Here, $k_x$= $2 \pi n /L_x$ and $k_y$= $2 \pi
m/L_y$, where $L_x$ and $L_y$ are the sides of our rectangular
simulation cell,    $n$ and $m$ being integers. To describe the gas-liquid
transition we used a cell of 73.79 $ \times$ 72.42 \AA$^2$, i.e., with a
surface 30 $\times$ 17 times that of the unit cell defined above.  
Introducing Eq. (\ref{Bloch}) into the Hamiltonian 
(\ref{hamiltonian}), the one-body Schr\"odinger equation transforms into
\begin{eqnarray} \label{Bloch2}
H u({\bf r_i})_{\bf k} &=&  \nonumber 
\left(\frac{\hbar^2}{2m} \left( \frac{1}{i} {\bf \nabla + k} \right)^2 + V_{ext}(x_i,y_i,z_i) \right) u({\bf r_i})_{\bf k} \\ 
 &=& E_{\bf k} u({\bf r_i})_{\bf k} 
\end{eqnarray}
We solved numerically this complex eigenvalue-eigenvector problem by
expanding 
\begin{eqnarray}
\lefteqn{u({\bf r_i})_{\bf k}  =   \frac{\sqrt 2} {\sqrt{(l_x l_y l_z)}}  
                       \sum_{j_1=-n_x}^{n_x} \sum_{j_2=-n_y}^{n_y} 
		       \sum_{j_3=1}^{n_z} c_{j_1 j_2 j_3} } \nonumber \\
		     & &  \times \exp[i( j_1 g_x x_i +  j_2 g_y y_i)] \sin(j_3 g_z (z_i-z_0)) 
\end{eqnarray}
and solving for the $c_{j_1 j_2 j_3}$ coefficients.
Here,  $n_x$ = 4, $n_y$=6, and $n_z$ = 30; $g_x$ = $2 \pi/l_x$, $g_y$ = $2
\pi/l_y$, and $g_x$ = $\pi/l_z$; $l_x$ = 2.46 \AA, $l_y$= 4.26 \AA \ and $l_z$
=  8-$z_0$ \AA$ $ ($z_0$ = 1.5 \AA). The solutions of Eq. (\ref{Bloch2})
were not restricted to be real. We used the number of functions necessary
to assure us an energy cutoff of 0.001 K. The ground state of a single
$^3$He atom obtained using this method was $E_{(0,0)}$ = $E_0$ = -135.771
$\pm$ 0.001 K. A plot of  $u(x,y,z=2.88)_{(0,0)}$ is displayed in Fig.
\ref{cut}, showing the corrugation of the ground state. That value of $z$
is the one for which the value of the wavefunction is maximum.  

Using the method described above, we obtained the $u({\bf r_i})_{\bf k}$
functions corresponding to the first band of the periodic potential created
by the graphite substrate. With them,  and using Eq. (\ref{Bloch}), we can
construct the one-body functions entering in the Slater determinants in Eq.
(\ref{trial1}). However, we found that, at least for the first-band functions we needed, all 
the $u({\bf r_i})_{\bf k}$ were 
real and independent of $\bf{k}$
within the
numerical errors derived from the procedure. This transforms Eq.
(\ref{trial1}) into   
\begin{equation}
\Phi({\bf r_1, r_2, ..., r_N}) =  D'^{\uparrow} D'^{\downarrow}  \prod_i u({\bf r_i}) 
\prod_{i<j} \exp \left[-\frac{1}{2}  
\left(\frac{b_{\rm He-He}}{r_{ij}} \right)^5 \right] 
\label{trial2}
\end{equation}
with $D'^{\uparrow}$ and $D'^{\downarrow}$ 
the plane-wave Slater determinants of the strictly 2D system
\cite{grau,bonin}. We also shifted the
coordinates entering these Slater determinants by introducing backflow
correlations in the standard way 
\begin{eqnarray}
\tilde x_i  & = & x_i + \lambda \sum_{j \ne i} \exp [-(r_{ij} -
r_b)^2/\omega^2] (x_i - x_j) \\ 
\tilde y_i  & = & y_i + \lambda \sum_{j \ne i} \exp [-(r_{ij} -
r_b)^2/\omega^2] (y_i - y_j) \ .
\end{eqnarray} 
We tested that the best parameters in those last equations were those
corresponding to the full three-dimensional homogeneous system
\cite{casulleras}, i.e., $\lambda = 0.35$; $\omega = 1.38$ \AA,  and $r_b
= 1.89$ \AA, instead of the ones corresponding to a pure 2D \cite{grau}.  
We made standard checks on the mean population of configurations and time
step to reduce any systematic bias to the level of the statistical noise.
Also, we included standard finite-size corrections to the energy coming
from both the discretization of the Fermi sphere and to the potential
energy contributions beyond the size of the simulation box.

\begin{figure}
\begin{center}
\includegraphics[width=0.7\linewidth]{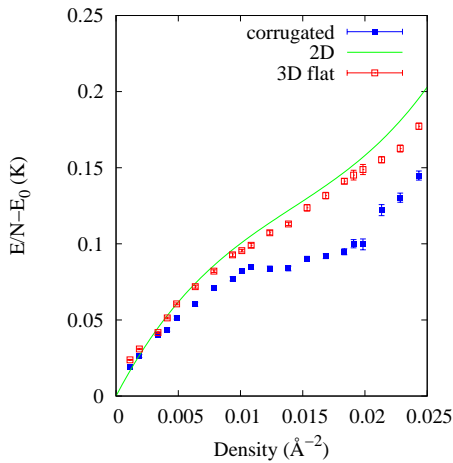} 
\caption{(Color online) Energy per $^3$He atom (E/N) versus surface density for corrugated 
(full symbols) and flat (open symbols) graphite. 
In both cases, we subtracted  the energy
in the infinite dilution limit, $E_0$, to allow for a better comparison.
The equation of state of a pure 2D liquid given in Ref. \onlinecite{grau} is also given 
for comparison.
}
\label{fig1}
\end{center}
\end{figure}

The results of the DMC simulations that consider a fully corrugated C-He
potential are
displayed in  Fig. \ref{fig1} as full squares. As indicated above, we
considered a simulation cell of 73.79 $ \times$ 72.42 \AA$^2$ including up
to 130 atoms, half of them with spin up, and the other half with spin
down.  In that figure, it is also included the strictly 2D results of Ref.
\onlinecite{grau} as a full line. To afford a comparison between the two
sets of data, in the first case we subtracted the energy in the  infinite
dilution limit ($E_0$) to the energy per particle ($E/N$). What we see is
that, apart from the limit when $\rho \rightarrow 0$, there is a sizeable
difference between both sets of data,  including an a
significant energy
stabilization for the full quasi-two-dimensional systems for $\rho > 0.005$
\AA$^{-2}$. 

To check if that decrease in the energy per
particle is due simply to the inclusion of the
additional  degree of freedom in the $z$ axis or there is something else,
we performed an additional FN-DMC calculation using an averaged-over version
of the external potential of Eq. (\ref{hamiltonian}) in the  $z$ axis.
Using a similar procedure to the one outlined above, 
we obtained $u(z_i)$, the   
one-body part of the trial function.
The open symbols in Fig. \ref{fig1} are the results of that
additional FN-DMC calculation. Here, as before, the energy in the infinite
dilution limit corresponding to the adsorption of a single $^3$He atom on
that flat graphite model was  subtracted. Its absolute value was slightly
smaller than 
that of the fully corrugated case ($E_0 = -133.585 \pm 0.001$ K). For that
second case, the energies per particle are much closer to the ones
corresponding to a pure 2D system. This is in line   
with the prediction of Ref. \cite{bonin}, but contradicts the results of 
Ref. \cite{brami}, where only smoothed-out substrates
were studied.

\begin{figure}
\begin{center}
\includegraphics[width=0.7\linewidth]{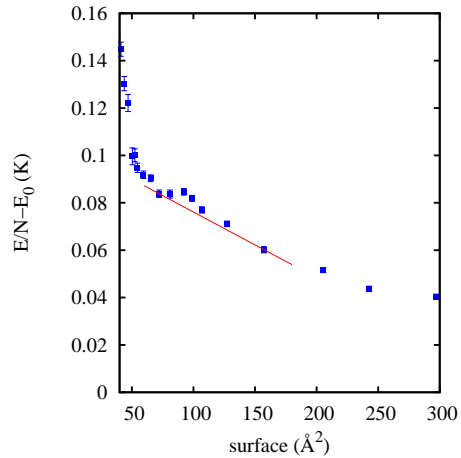} 
\caption{(Color online) Full corrugated results for the energy per $^3$He atom as a 
function of the surface per particle. The line stands for the
double-tangent Maxwell construction to determine the equilibrium densities
of the gas and liquid phases.  
}
\label{fig2}
\end{center}
\end{figure}

In Fig. \ref{fig2}, we show the same
corrugated data as in Fig. \ref{fig1},
but as a function of the inverse of the $^3$He density.  At first sight, we
can see that there is a non-stability zone  around a surface per particle
of around 100 \AA$^2$. In that figure it is also displayed the double-tangent
Maxwell construction line (see Ref. \cite{chandler} for details about its construction).
This allows us to see that there is indeed a
first-order phase  transition between a dilute gas of density $0.006 \pm
0.002$ \AA$^{-2}$, and a liquid one of $0.014  \pm
0.002$ \AA$^{-2}$. We can
assign tentatively that transition to the gas-liquid  equilibrium suggested
in Ref. \cite{fukuyama} for $^3$He on clean graphite. We have to
stress also that we did not use different trial wavefunctions for gas and
liquid phases, the instability appearing naturally  when we increase the
helium density.   

\begin{figure}
\begin{center}
\includegraphics[width=0.7\linewidth]{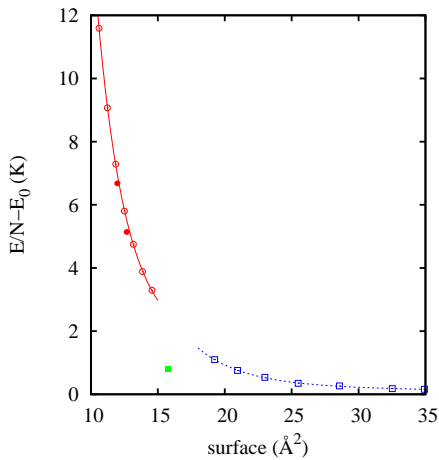} 
\caption{(Color online) Same as for Fig \ref{fig2} but for higher values of
the surface per particle.  Open squares, liquid phase; open circles, incommensurate solid; full square, 
$\sqrt3 \times \sqrt3$  phase; full circles, $31/75$ (lower one) and $7/16$ (upper one), 
registered phases. The error bars are of the size of the symbols and not displayed for simplicity. The guides are mere guides-to-the-eye.
}
\label{fig3}
\end{center}
\end{figure}

If we increase further the amount of helium adsorbed, the system will
undergo another phase transition, in this case to a $\sqrt3 \times \sqrt3$
registered phase, similar to that of $^4$He on graphite.    This is
illustrated in Fig. \ref{fig3}. There, we plot the energy per particle
for a liquid (open squares), an
incommensurate solid (open circles), and several registered structures
(full symbols). The calculations for the liquid phase, quite beyond the
transition point, were done using the
same procedure outlined above but using a smaller simulation cell, one with
a surface of 44.28 $\times$ 42.6 \AA$^{-2}$. To model the solid structures,
we followed Ref. \onlinecite{yo}, and multiplied the trial function of Eq.
(\ref{trial1}) by a Nosanow factor, 
\begin{equation}
\prod_i \exp \{-a [(x_i-x_{\rm site})^2 + (y_i-y_{\rm site})^2] \} \ ,
\label{trial3}
\end{equation}
where $x_{\rm site},y_{\rm site}$ are the coordinates of the
crystallographic positions of the solid structures, and $a$ was
variationally optimized ($a$=0.24 \AA$^{-2}$ for all
the lattices). To establish
 the boundaries between the liquid phase and  the
$\sqrt3 \times \sqrt3$ commensurate structure, we would have to do another
double-tangent Maxwell construction.  We proceeded in the same way as in
previous literature, by drawing the line with the smallest negative slope
that goes from the inverse of the solid density and intercepts the liquid
equation of state.  We found that the  $\sqrt3 \times \sqrt3$ solid is in
equilibrium with a liquid of density 0.039 $\pm$ 0.001 \AA$^{-2}$, i.e.,
the stability range of the liquid is from   0.014  $\pm$ 0.002 \AA$^{-2}$
to 0.039 $\pm$ 0.001 \AA$^{-2}$. The latter value is in good agreement with
the experimental upper value for a liquid phase found for a three-layer
$^3$He system \cite{godfrin}. 
The smallest value of the interval is compatible with the experimental
lower value for the same system \cite{fukuyama1}.   

The full symbols in Fig. \ref{fig3} correspond to two commensurate solids
already considered for quantum species on graphite, the $31/75$ ($\rho$ =
0.0789 \AA$^{-2}$, found for D$_2$ \cite{frei,yo4}),  and the $7/16$
($\rho$ = 0.0835 \AA$^{-2}$, proposed to be stable by Corboz
\textit{et al.} for
$^4$He \cite{corboz}). As one can see in that figure, we found that those
registered solids are slightly more stable than  the corresponding incommensurate
(IC) solids of the same density. In particular, the energies per atoms are: 
$E_{31/75}$ = -130.63 $\pm$ 0.02 K versus $E_{\text{IC}}$ = -130.30 $\pm$
0.01 K, and $E_{7/16}$ = -129.09 $\pm$ 0.02 K versus $E_{\text{IC}}$ =
-129.00 $\pm$ 0.02 K. This means that if we increase the density beyond the
one corresponding to the $\sqrt3 \times \sqrt3$ structure (0.0636
\AA$^{-2}$), the system will undergo a first-order phase transition to a 
registered $31/75$ structure that, on further increase will transform into
a $7/16$ one. Obviously, these latter phase transitions are predicted to exist in the limit
of zero temperature and they could be smoothed out if the temperature is not
low enough due to the small energy differences here obtained. At
higher densities, there is a last 
transition into an incommensurate
triangular solid. Another Maxwell construction  using the data shown in
Fig. \ref{fig3} allowed us to obtain that the lower density limit
of this phase is 0.089 $\pm$ 0.005 \AA$^{-2}$. 

The results presented allow us to give a coherent picture that can
incorporate all the experimental results on $^3$He on 
graphite. The very dilute density for the liquid phase found  in Refs.
\cite{gasparini,fukuyama} ($\sim$ 0.006 \AA$^{-2}$) is compatible
with  our lower limit for the gas-liquid transition. This means that
for ranges 0.006 \AA$^{-2} < \rho <$ 0.014 \AA$^{-2}$ the system will
separate itself into a very dilute gas phase and puddles of liquid of density 0.014
\AA$^{-2}$, in the right proportions to produce the density we considered
within that interval.  So, from 0.006 \AA$^{-2}$ up, we will have part of
the surface covered by a liquid. That coverage will be complete when the
overall $^3$He density is $\rho$ = 0.014 \AA$^{-2}$, in which all graphite
will be coated by an homogeneous liquid. That liquid will be stable up to
0.039 $\pm$ 0.001 \AA$^{-2}$, in line with the results of Ref.
\onlinecite{godfrin}. 
On the other hand,  we see that $^3$He presents two
new stable registered phases at relatively high densities. The only
experimental support for the first one ($31/75$) are the calorimetric
measurements of  Greywall \cite{greywall}, in with a  $^3$He first-layer
solid phase on graphite at $\rho$ = 0.076 \AA$^{-2}$ is considered. However,
the phase proposed is a $2/5$ one, that we found to be unstable with 
respect to an incommensurate triangular solid of the same density. 
Our results show that the main, and forgot up to now, ingredient to
satisfactorily describe the monolayer of $^3$He on graphite is the use of a
realistic C-He interaction instead of smoothed or averaged
surface-helium potentials.

\acknowledgments
We acknowledge partial financial support from the 
MINECO (Spain) 
grants No. FIS2014-56257-C2-2-P and No.FIS 2014-56257-C2-1-P, and 
Junta de Andaluc\'{\i}a group PAI-205 and grant FQM-5987.

%\bibliographystyle{unsrt}
%\bibliography{refs}

\end{document}